\documentclass{article}

\usepackage{PRIMEarxiv}

\usepackage[utf8]{inputenc} 
\usepackage[T1]{fontenc}    
\usepackage{hyperref}       
\usepackage{url}            
\usepackage{booktabs}       
\usepackage{amsfonts}       
\usepackage{nicefrac}       
\usepackage{microtype}      
\usepackage{lipsum}
\usepackage{fancyhdr}       
\usepackage{graphicx}       
\graphicspath{{media/}}     

\title{\textnormal{Exploring the Potential of AI-Generated Synthetic Datasets: A Case Study on Telematics Data with ChatGPT}}

\author{
  Ryan Lingo \\
  Honda Research Institute, USA\\
  \texttt{ryan\_lingo@honda-ri.com} \\
}

\begin{document}
\maketitle

\begin{abstract}
This research delves into the construction and utilization of synthetic datasets, specifically within the telematics sphere, leveraging OpenAI's powerful language model, ChatGPT. Synthetic datasets present an effective solution to challenges pertaining to data privacy, scarcity, and control over variables - characteristics that make them particularly valuable for research pursuits. The utility of these datasets, however, largely depends on their quality, measured through the lenses of diversity, relevance, and coherence. To illustrate this data creation process, a hands-on case study is conducted, focusing on the generation of a synthetic telematics dataset. The experiment involved an iterative guidance of ChatGPT, progressively refining prompts and culminating in the creation of a comprehensive dataset for a hypothetical urban planning scenario in Columbus, Ohio. Upon generation, the synthetic dataset was subjected to an evaluation, focusing on the previously identified quality parameters and employing descriptive statistics and visualization techniques for a thorough analysis. Despite synthetic datasets not serving as perfect replacements for actual world data, their potential in specific use-cases, when executed with precision, is significant. This research underscores the potential of AI models like ChatGPT in enhancing data availability for complex sectors like telematics, thus paving the way for a myriad of new research opportunities.
\end{abstract}


\section{Introduction}
The advent of Artificial Intelligence (AI) and Machine Learning (ML) has revolutionized numerous fields, expanding the realm of possibilities with vast computational abilities and innovative techniques. A critical element fueling this advancement is the accessibility of large-scale datasets. However, data availability often confronts challenges such as privacy concerns, scarcity of information, and constraints in manipulating variables for research. In response, the creation of synthetic datasets has emerged as a promising solution. Synthetic datasets, produced artificially, circumvent many of these issues while preserving the statistical properties of the original data. This allows for experimental control and confidentiality, making them particularly beneficial in data-sensitive fields like healthcare, finance, and the automotive industry.

While the potential of synthetic datasets is significant, their effectiveness is fundamentally dependent on their quality, a characteristic primarily defined by three aspects: diversity, relevance, and coherence. Ensuring high standards in these aspects is of utmost importance for synthetic data to be a reliable tool in research and applications.

This paper employs OpenAI's language model, ChatGPT, as a tool for generating synthetic data, with a focus on the telematics domain. The telematics sector, encompassing telecommunications and vehicular technologies, stands to benefit substantially from an increased availability of data for research, planning, and technological advancement. This paper showcases the procedure of crafting synthetic telematics data using ChatGPT through a case study, delving into the complications and corresponding solutions tied to this process, and assessing the quality of the resulting dataset. The intention lies in offering a glimpse into the potential of AI-generated synthetic datasets, underscoring their promise as a vital resource in the realm of data science.

\section{Related Work}
The emergent utility of synthetic data in recent years, owing to its potential in enhancing privacy and representation, has sparked research interest in various sectors. This section synthesizes the most significant studies in this field, focusing particularly on the application of synthetic data and the role of artificial intelligence.

Savage [2] pioneers the discourse in the paper "Synthetic data could be better than real data", asserting that machine-generated data sets could potentially augment both privacy and representation in artificial intelligence. However, Savage cautions that the successful application of synthetic data hinges on the delicate balance between achieving accurate representation and mitigating privacy invasion.

Complementing Savage's narrative, Stadler et al. [3] introduce an empirical perspective in their paper "Synthetic Data – Anonymisation Groundhog Day." The authors execute a quantitative evaluation of the privacy gain associated with synthetic data publishing. In a stark comparison with preceding anonymisation techniques, they assert that synthetic data either inadequately thwarts inference attacks or it fails to retain data utility, indicating the need for a balanced approach.

Simultaneously, the application of synthetic data in healthcare is aptly highlighted by Gonzales et al. [1] in their paper "Synthetic data in health care: A narrative review." They identify seven potential use cases of synthetic data in healthcare, which encompass research, hypothesis testing, public health research, health IT development, education, training, public data release, and data linking.

Hassani and Silva [4] diversify this conversation by examining the implications of artificial intelligence, particularly ChatGPT, in the realm of data science. In their paper "The Role of ChatGPT in Data Science: How AI-Assisted Conversational Interfaces Are Revolutionizing the Field", they elucidate how ChatGPT can augment various aspects of the data science workflow. They contend that it can aid in data cleaning, preprocessing, model training, and result interpretation, and provide fresh insights to drive decision-making.

Despite the research surrounding synthetic data and ChatGPT, there exists a conspicuous knowledge gap regarding the use of ChatGPT in creating synthetic telematics data. The current body of literature hence necessitates further exploration into the prospective benefits and challenges of employing ChatGPT for synthetic telematics data generation.

\section{Theoretical Background and Potential of Synthetic Datasets}
\subsection{Introduction to Synthetic Data: A Paradigm Shift}
The digital revolution and the rise of data-centric decision making has catalyzed the emergence of a potent instrument—synthetic data. This data, algorithmically crafted without real-world collection, has gained prominence due to an escalating need for extensive, adaptable data that adheres to privacy and accessibility constraints. This study investigates the concept of synthetic data, exploring its creation, applications, potential shortcomings, and its utility as a powerful resource for beginners in data analysis and data science.

\subsection{Exploring the Characteristics of Synthetic Datasets}
In essence, a synthetic dataset is a conglomerate of artificially devised data points that ideally retain the statistical properties of a corresponding real-world dataset. This type of data emulates the features and behaviors of actual data without encompassing any confidential or sensitive information. The generation of synthetic data deploys intricate algorithms and techniques, including a range of machine learning methodologies, to assure that the generated data faithfully mirrors the attributes of the original dataset.

The ideal synthetic dataset demands meticulous replication of inherent patterns present in the original data. However, the dataset generated with ChatGPT might not conform to such rigorous standards. Considering the data is generated from randomized sources, there is an inevitable compromise in the depth and breadth of information compared to a comprehensive synthetic dataset. Critical to note is the primary objective - not to mirror reality flawlessly, but rather to enhance research capabilities and nurture educational ventures. The generated synthetic dataset in this paper, though perhaps not as exhaustive, provides a controlled environment. This environment allows researchers and learners to conduct experiments, explore, and derive insights, all the while ensuring privacy and confidentiality. In the pursuit of knowledge, such a dataset proves adept in facilitating understanding and analysis of intricate data structures and patterns.

\section{The Implications and Applications of Synthetic Datasets}
\subsection{A Panorama of Applications for Synthetic Datasets}
Synthetic datasets permeate a multitude of disciplines, with the following areas being notably dominant:
\begin{itemize}
\item \textbf{Machine Learning and Artificial Intelligence:} Synthetic data can prove instrumental in training machine learning models, particularly when the procurement of real-world data is restricted or infeasible due to privacy considerations.
\item \textbf{Testing and Validation:} Synthetic data is frequently employed in software testing to assess system performance across varying scenarios, ensuring no confidential data is disclosed during the process.
\item \textbf{Research and Development:} When accessing real-world data is challenging or ethically dubious, researchers often depend on synthetic data for experimental design and hypothesis testing.
\item \textbf{Data Augmentation:} Synthetic data can rectify dataset imbalance by generating data for underrepresented classes, thereby enhancing machine learning model performance.
\item \textbf{Privacy-Preserving Data Sharing:} In sectors dealing with sensitive data, such as healthcare, finance, and automotive industries, synthetic datasets facilitate secure data sharing while preserving privacy.
\item \textbf{Education and Learning:} Synthetic data can play a pivotal role in education, particularly for experiential learning in data science and machine learning. These datasets can be tailored to demonstrate specific concepts or techniques, negating the need for exhaustive data collection or dealing with privacy issues. This allows students, educators, and competition participants to experiment, make mistakes, and learn, all without jeopardizing real-world sensitive data.
\end{itemize}

Thus, synthetic datasets play a vital role in our progressively data-driven world, striking a balance between utility, privacy, and accessibility. Given the emphasis on the educational applications of synthetic data, it becomes highly relevant for the objectives of this study.

\subsection{The Role of Synthetic Data in Real-World Scenario Modelling}
Synthetic data is invaluable when modeling various scenarios. For instance, in telematics, which includes data collection about a vehicle's usage, location, and driver behavior, acquiring real-world data for every possible driving scenario can be arduous, if not impossible. Synthetic data can simulate a broad spectrum of conditions, from normal to extreme, enabling comprehensive testing and training of telematics systems, contributing to the evolution of safer, more efficient vehicle technologies.

\subsection{Synthetic Data: An Ally for Data Privacy}
In our digital era, data privacy is of paramount concern. Synthetic data emerges as a privacy champion, allowing organizations to draw insights from data without disclosing sensitive or confidential information. For instance, in telematics, the data collected often includes sensitive data, such as precise location data and potentially personally identifiable information (PII). By generating synthetic datasets that maintain the statistical properties of the original without including any actual PII, privacy can be preserved during in-depth analyses, predictive model development, or data sharing with third parties. It's also worth noting that in scenarios where achieving a synthetic dataset that is statistically identical to the original proves unfeasible, a dataset that is statistically close enough can still serve effectively for educational and experimental applications. This slightly less precise alternative yet maintains essential characteristics for exploration and understanding, offering a robust solution for privacy-preserving data utilization.

\subsection{Synthetic Data and its Impact on Scalability and Flexibility}
Algorithmically generated synthetic data provides scalability and flexibility, allowing organizations to produce as much (or as little) data as necessary for their distinct applications. This aspect is crucial when stress-testing systems or training intricate machine learning models. The versatility of synthetic data enables its adaptation to represent specific scenarios, populations, or conditions, thereby enabling more targeted and precise analyses and predictions. It is noteworthy to add that the specific nature of data and its level of congruence with the original source will vary from one use case to another. This reflects the bespoke nature of synthetic data generation, making it a vital tool for a wide range of scenarios. The unprecedented ease in creating synthetic data today necessitates further scholarly exploration of its various potential applications. Hence, the need for continued research to decipher the maximum potential of synthetic data in all possible use cases remains integral to advancing our knowledge in this rapidly evolving field.

\subsection{Addressing Data Imbalance with Synthetic Data}
Imbalanced data poses a significant challenge in machine learning, often leading to models biased towards overrepresented classes. Synthetic data offers a resolution by generating additional data for the underrepresented classes, thus creating a more balanced dataset that improves learning and facilitates fairer predictions. For instance, in telematics, certain infrequent driving behaviors can be artificially generated, producing a more balanced dataset for model training and leading to advancements in driver safety and vehicle performance. Building upon this notion, not just the creation, but also the augmentation of synthetic data using large language models (LLMs) could emerge as one of the most crucial use cases. When a sample of original data is available, it is feasible to map its distributions onto the synthetic data, which enhances the authenticity and richness of the synthetic dataset. This specific application, which encompasses both creation and augmentation of synthetic data with LLMs, represents a crucial frontier in machine learning. Given its immense potential to reshape our approach to data imbalances, it justifiably necessitates extensive, focused research.

\section{The Inherent Challenges of Utilizing Synthetic Data}
\subsection{Limitations in the Complexity and Variability of Synthetic Data}
While synthetic data provides a multitude of possible benefits, it is not without its inherent challenges. A primary concern is the potential lack of full complexity and variability that characterizes real-world data. Being algorithmically generated, synthetic data might fail to fully encapsulate the intricate details, correlations, and unpredictability found in its real-world counterparts. This shortfall may result in models that perform well with synthetic data but do not generalize effectively to real-world scenarios.

To illustrate this point, consider telematics. A synthetic dataset might simulate a range of driving conditions, but it might not account for subtle, influential factors such as driver fatigue, distractions, or even the nuanced effects of billboard advertisements along a route. These omissions could significantly impact the performance of predictive models when applied to real-world data.

\subsection{The Risk of Inherited Biases in Synthetic Data}
Another inherent challenge of synthetic data is its potential to propagate the biases embedded in the model used for its generation. If the model creating the synthetic data contains biases, whether from the original training data or algorithmic bias, these biases can be transferred to the synthetic data, leading to skewed results and potentially biased conclusions.

For instance, suppose a model generating synthetic telematics data was primarily trained on urban driving conditions. In that case, it might inadvertently over-represent these conditions in the synthetic data, leading to an under-representation of rural or off-road driving scenarios. Such biases could potentially skew the conclusions drawn from the analysis or predictions based on this data, thereby affecting the validity of decisions made using these insights.

\subsection{Overfitting Risk Associated with Synthetic Data}
The exclusive use of synthetic data for model training can lead to overfitting—a situation where a model exhibits high performance on the training data but struggles to generalize to new, unseen data. This issue can arise if the synthetic data does not adequately represent the variability inherent in real-world scenarios.

In the domain of telematics, consider a model trained on synthetic data to predict engine failure based on various parameters, such as driving style, vehicle load, and environmental conditions. If the synthetic data does not capture the full spectrum of real-world variations and complexities, the model might become excessively specialized to the synthetic dataset, leading to overfitting. Consequently, the model's predictive performance may decline when applied to real-world data

\subsection{Addressing the Challenges of Synthetic Data}
The objections raised towards synthetic data, although reasonable, are not insurmountable. With strategic advancements and diligent application, these concerns can be reduced.

The challenge of fully capturing the complexity and variability of real-world data can be approached by refining the data generation algorithms and including a wider set of influencing factors and conditions. In the context of telematics, this could mean extending the generation process to factor in additional elements such as driver fatigue or distractions. While this introduces an added layer of complexity, it holds the potential for enhancing the practicality and precision of synthetic data.

The potential risk of inherited biases in synthetic data can be mitigated through balanced and careful training of the generative models. This involves incorporating diverse and representative datasets during the training phase to prevent over- or under-representation of specific conditions or classes. Continuous monitoring and correction of any observable bias in the generated synthetic data through iterative improvements in the data generation process is also essential.

As for the risk of overfitting associated with synthetic data, a mixed approach utilizing both real and synthetic data for model training could be adopted. This hybrid strategy could leverage the benefits of synthetic data while preserving the capacity to generalize to real-world conditions. Techniques such as regularization, cross-validation, and ensemble methods can also be deployed to counter overfitting. For instance, in the telematics example, a blend of real-world data and synthetic data encompassing various driving conditions could lead to a more robust and generalizable model.

To summarize, while synthetic data indeed presents challenges, these are not insurmountable barriers. It's essential to underscore that while all these concerns are valid and warrant attention, the scope of this paper is primarily to initiate the conversation around generating a telematics dataset using ChatGPT. With careful application, continuous improvement, and strategic advancements, synthetic data can prove an invaluable tool in data science, particularly in data-sensitive fields, and further research in this area is merited

\section{An Examination of ChatGPT's Role in Synthetic Dataset Generation}
\subsection{Leveraging ChatGPT in Language Modelling for Synthetic Data Generation}
ChatGPT, a product of OpenAI, represents an advanced language model utilizing machine learning to generate text reminiscent of human discourse. Its potential extends beyond mere text generation, enabling the creation of structured data in response to specific prompts.

Leveraging the capabilities of ChatGPT, this paper proposes a practical demonstration involving the generation of a synthetic dataset. The focus will be on crafting a rudimentary telematics dataset. This dataset, while simplified, is intended to serve as an educational resource, encapsulating key data points inherent in telematics analytics.

This study emphasizes the creation of a synthetic dataset with composition limited to randomly generated values. While this approach may not replicate the direct correspondence found in real-world data, it serves as a highly conducive environment for exploratory research, analytical processing, and in-depth learning. In fact, the apparent simplicity of this randomly generated data allows for clearer understanding and manipulation, providing an ideal introductory tool in the realm of telematics data.

Throughout the process of generating and exploring this synthetic dataset, it is critical to bear in mind the primary objective is not an exact emulation of real-world telematics data. Instead, the focus lies in providing an accessible gateway for individuals aiming to develop a foundational understanding of such complex data. Thus, the value of this synthetic dataset lies not in its replication of real-world data, but in its ability to facilitate learning, discovery, and comprehension of telematics data structure and interpretation.

\subsection{Methodological Considerations: Utilizing ChatGPT for Synthetic Dataset Creation}
In the generation of synthetic datasets using ChatGPT, the initial stage involves the delineation of the data's anticipated format and structure. When contemplating a scenario involving tabular traffic pattern data, for instance, the structure might comprise elements such as date, time, location, traffic volume, and weather conditions.

Following this, the process segues into the formulation of a sequence of prompts which serve to guide ChatGPT about the character of the data intended for generation. Armed with these well-structured prompts, ChatGPT stands primed to assist effectively in the orchestration of a synthetic telematics dataset.

\subsection{Extending Synthetic Data Applications Beyond Demonstrative Purposes}
Engaging with the process of synthetic data creation entails acknowledging its extensive potential that extends beyond mere research and demonstration. In scenarios where synthetic data are expected to integrate into production or operational environments, the process is not restricted to the act of data generation alone.

A vital part of the process is the incorporation of a robust validation and testing phase, essential for validating the quality of the synthetic data and evaluating its suitability for its intended function. This crucial phase could involve statistical comparisons with real data and the employment of synthetic data within various models.

Though this paper primarily concentrates on delving into potential applications, rather than preparing for real-world operational use, the importance of a comprehensive validation process should not be diminished for those intending to broaden their exploration of synthetic data. This pivotal step facilitates the unlocking of a synthetic dataset's full potential, thus charting the path towards robust, privacy-conscious, and holistic data solutions.

\section{An Exploration of Synthetic Telematics Data Generation using ChatGPT}
\subsection{Introduction to Telematics}
Telematics, the synergistic blend of telecommunications and informatics, plays a pivotal role in today's data-driven era, specifically within the realms of vehicular operation and urban mobility. At its core, telematics is concerned with the collection, processing, and transmission of information pertaining to remote entities, predominantly vehicles, facilitated by telecommunication devices.

Consequently, the breadth of telematics data encapsulates numerous informational parameters, including a vehicle's precise location, speed, idle time, instances of harsh acceleration or abrupt braking, fuel consumption, along with other intricate diagnostic data. This wealth of information is typically harvested via a combination of GPS technology, onboard diagnostics, and sensory devices. The richness and scope of telematics data make it a compelling field for the application of synthetic data generation.

\subsection{The Influence of Telematics Data}
Telematics data holds the power to deliver transformative insights across diverse sectors, with particular prominence in urban planning and traffic management. A comprehensive understanding of traffic patterns, derived from this data, is an indispensable asset for managing road infrastructure planning and traffic congestion. This knowledge facilitates the pinpointing of peak congestion times and zones, the formulation of strategies for optimizing traffic flow, the planning of enhanced public transportation routes, and the overall improvement of the commuter experience.

With the growing ubiquity of Internet of Things (IoT) devices and the enhancement of connectivity, telematics data has emerged as a critical component in smart city initiatives. Real-time traffic data can be harnessed to steer emergency services towards quicker response times, or to inform eco-routing strategies aimed at curtailing carbon emissions. Therefore, the potential and impact of telematics data extend far beyond its conventional applications, leading the charge towards a smarter and more sustainable urban environment.

\subsection{Telematics Data in Research and Innovation}
Beyond immediate operational applications, telematics data also has immense potential for research and innovation. Researchers can leverage this data to predict future traffic trends, improve urban planning, and facilitate the development of autonomous vehicles. Given the growing concern for privacy and data security, synthetic telematics data, mirroring real-world data without compromising privacy, has become a significant research area. Synthetic datasets can aid researchers in bypassing privacy-related obstacles and uncovering new insights in urban mobility.

\subsection{Generation of Synthetic Telematics Data using ChatGPT}
The process of synthesizing a telematics dataset with ChatGPT involves prompt experimentation and iterative refinement, with gradual increases in prompt complexity. The initial step starts with a simple open-ended prompt such as, “Can you create a telematics dataset for me?” This step provides an insight into ChatGPT's interpretation and structuring of telematics data.


\begin{figure}[h]
\begin{center}
\includegraphics[width=80mm]{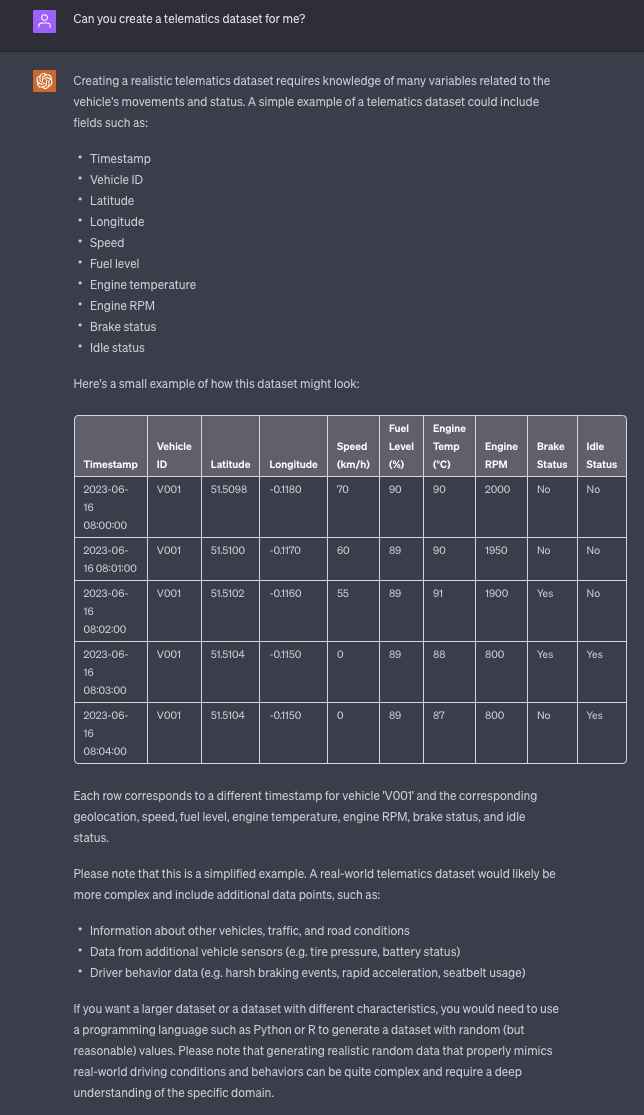}
\caption{Initial output from ChatGPT in response to the open-ended prompt: 'Can you create a telematics dataset for me?' - illustrating the AI's interpretation and structuring of telematics data.}
\end{center}
\end{figure}

This initial, basic prompt serves as our first insight into how ChatGPT understands and structures telematics data. It forms the foundation upon which we build, enhancing our prompts iteratively for better and more specific outputs.

The beauty of this approach lies in its progressive nature, reminiscent of a teaching or training process where we start with the basics and gradually introduce more complex elements.

From our first prompt, "Can you create a telematics dataset for me?", we glean several key observations. Firstly, ChatGPT demonstrates an understanding of what telematics data is, signifying that it is capable of generating data relevant to our use case.

However, the output is quite short, consisting of only five rows, which indicates that we may need to provide clearer instructions about the size of the dataset we desire.

Moreover, while the model generates telematics data, the output format is not readily usable. The data is not in a format, such as a CSV or Pandas' DataFrame, that is amenable to further analysis or processing.

These crucial insights inform us that while ChatGPT understands the concept of a 'telematics dataset', it requires more specific instructions to generate the type and format of data we're targeting. Consequently, these revelations guide us as we craft more complex prompts in our subsequent iterations.

Moving forward in our data generation journey, our second prompt brings more specificity to the table, directly addressing some limitations we noticed with our initial, more open-ended prompt. Our refined prompt reads, "Can you create a telematics dataset for me that has 100,000 rows and is exportable into CSV format or a Pandas DataFrame?"


\begin{figure}[h]
\begin{center}
\includegraphics[width=80mm]{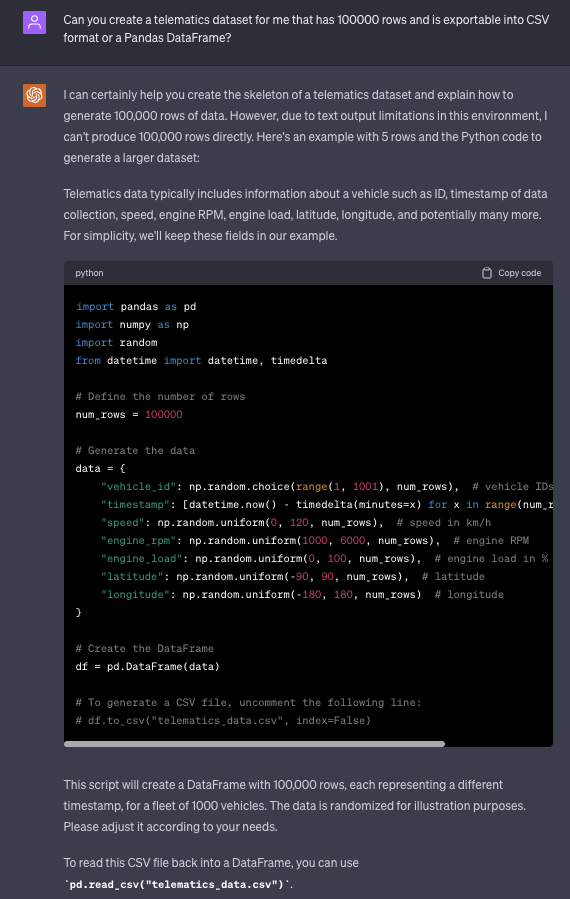}
\caption{ChatGPT's response to the refined prompt: 'Can you create a telematics dataset for me that has 100,000 rows and is exportable into CSV format or a Pandas DataFrame?' - showcasing the impact of increased specificity on synthetic data generation.}
\end{center}
\end{figure}

This expanded prompt offers a few key enhancements:
\begin{enumerate}
  \item \textbf{Dataset Size:} Explicitly requesting 100,000 rows sets a clear expectation for the scale of the dataset, influencing potential data generation methods and potential optimizations for handling larger datasets.
  \item \textbf{Format:} The prompt's specification for an exportable CSV format or a Pandas' DataFrame offers clear guidance on the dataset's structure.
\end{enumerate}

The second attempt underscores the value of providing explicit instructions in the prompts. Precise requirements can minimize the likelihood of needing multiple iterations and can enhance the chances of producing a synthetic dataset aligned with research needs from the outset.

Although the second prompt exhibits marked improvement over the initial attempt, the refinement process is iterative, offering continual opportunities for enhancement. The objective is not to produce a flawless synthetic telematics dataset at the first instance, but to understand the capabilities and limitations of the ChatGPT model more deeply, and to learn how to steer it towards producing the needed data.

With this insight, the focus shifts to the creation of the third prompt. Building upon the improvements achieved so far, the objective becomes to strike an optimal balance between specificity and data diversity. This approach aims to more effectively guide ChatGPT in the generation of a synthetic dataset that is both usable and better aligned with the requirements.

Venturing further into the exploration of synthetic dataset creation, the third prompt incorporates an additional layer of complexity: context. This step involves asking ChatGPT to envision a real-life scenario, providing a more tangible backdrop for the data generation process.

The updated prompt is:
"Imagine you are a city planner in Columbus, Ohio. You have been tasked with studying the traffic patterns around the city. The goal is to help people who have never worked with telematics data have a synthetic dataset they could practice on. Can you create a telematics dataset that has 100,000 rows and is exportable into CSV format or a Pandas DataFrame? I would like it to have the following columns: ['driver id', 'timestamp of trip start time', 'timestamp of trip end time', 'longitude of trip start', 'latitude of trip start', 'longitude of the trip end', 'latitude of the trip end', 'average speed', 'day of the week', 'vehicle type', 'heading', 'road type', 'weather conditions'], as well as any other columns you think would be useful."

Introducing not only a contextual backdrop to guide data generation, but also explicitly requesting specific columns previously unmentioned, adds further depth to the synthetic data generation. This third prompt places ChatGPT in the shoes of a city planner in Columbus, Ohio, intending to generate a synthetic telematics dataset. Simultaneously, it encourages the model to draw from its vast training data and suggest additional variables that might enhance analysis, by inviting it to introduce any other columns it deems useful.

This example underscores the iterative nature of refining AI requests. By providing a context, explicitly defining the size and format of the dataset, and leaving room for the model's creativity, it edges closer to a synthetic dataset that meets research and learning requirements. The outcome and insights from this third endeavor provide the next topic of exploration.

\subsection{Dealing with Challenges and Debugging}
Diving deeper into the complexity of prompts naturally results in more elaborate outputs, and consequently, the likelihood of encountering issues elevates. Nonetheless, it's essential to view these problems not as setbacks but integral parts of the process. They present opportunities to better understand the workings of ChatGPT and, importantly, how to leverage its abilities to address the issues that surface.

During the third iteration, the first coding issue emerged. The code produced by ChatGPT did not execute as anticipated and returned an error. This occurrence is not uncommon when working with AI models such as ChatGPT. By escalating the complexity of the task, the likelihood of encountering an issue proportionately increases. However, this is precisely where the advantage of using AI becomes apparent.

Instead of manually debugging the issue, ChatGPT was utilized to identify the problem and generate a solution. Simply pasting the error back into ChatGPT elucidated the nature of the error and yielded rectified code that executed without glitches.

This real-world scenario illustrates the adaptability and power of ChatGPT. Its capability to understand and correct its own errors further emphasizes its potential in generating increasingly complex synthetic datasets.


\begin{figure}[h]
\begin{center}
\includegraphics[width=100mm]{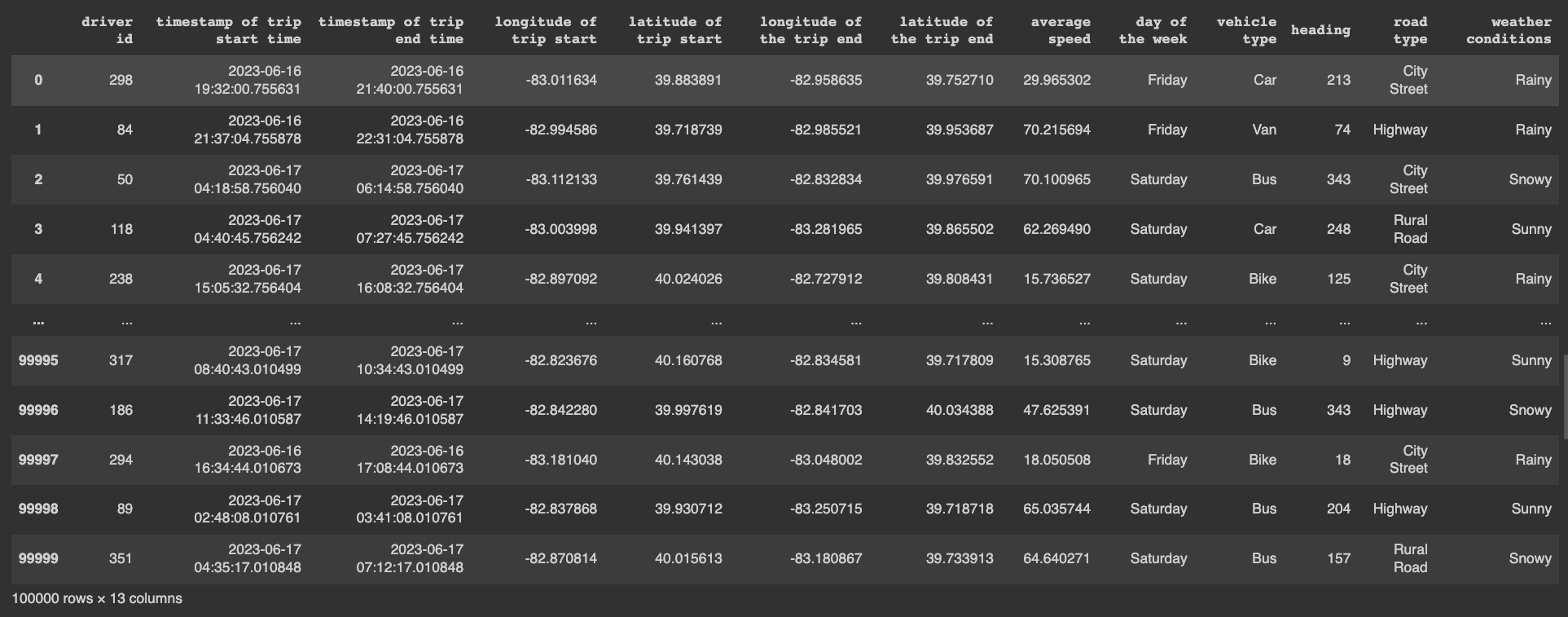}
\caption{Snapshot of the synthetic telematics dataset as viewed in a Pandas DataFrame.}
\end{center}
\end{figure}

\section{A Methodical Approach to Evaluating the Quality of a Synthetic Dataset}
Evaluating the quality of a synthetic dataset calls for a meticulous, systematic approach, especially when no real-world comparison dataset is available. Using the created synthetic telematics dataset as an example, the following stages are suggested to effectively assess its quality.

\subsection{Diversity Examination}
The dataset should exhibit a reasonable range across all variables. For example, a dataset in which all trips are of the same length, or where a single type of weather condition prevails, may not provide insightful results. A prompt for generating Python code to examine the data distribution might be as follows: “Show how to generate summary statistics for each column in a Pandas DataFrame for understanding the data distributions within a telematics dataset.” Keep in mind that the name of the CSV may require modification to align with the one used during dataset creation.


\subsection{Relevance Evaluation}
Data should align with the real-world context it is intended to reflect. Therefore, it becomes necessary to ensure that vehicle speeds fall within city-appropriate limits and that weather conditions are congruent with the climate typical to Columbus, Ohio. A Python code prompt for checking such relevance could be: “Provide a method to visualize the distribution of average speeds in a synthetic telematics dataset. Could you generate a code snippet for a histogram?” The output following this relevance examination is provided in the subsequent section.

\begin{figure}[h]
    \centering
    \begin{minipage}{0.45\textwidth}
        \centering
        \includegraphics[width=1\linewidth]{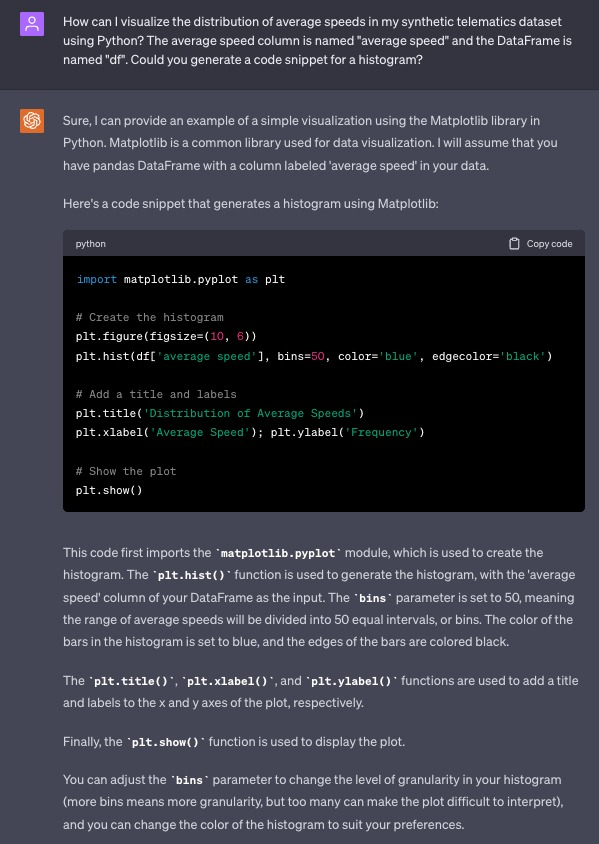} 
        \caption{Output of ChatGPT when queried for generating a Python code snippet to visualize the distribution of average speeds in the synthetic telematics dataset.}
        \label{fig:prompt5}
    \end{minipage}\hfill
    \begin{minipage}{0.45\textwidth}
        \centering
        \includegraphics[width=1\linewidth]{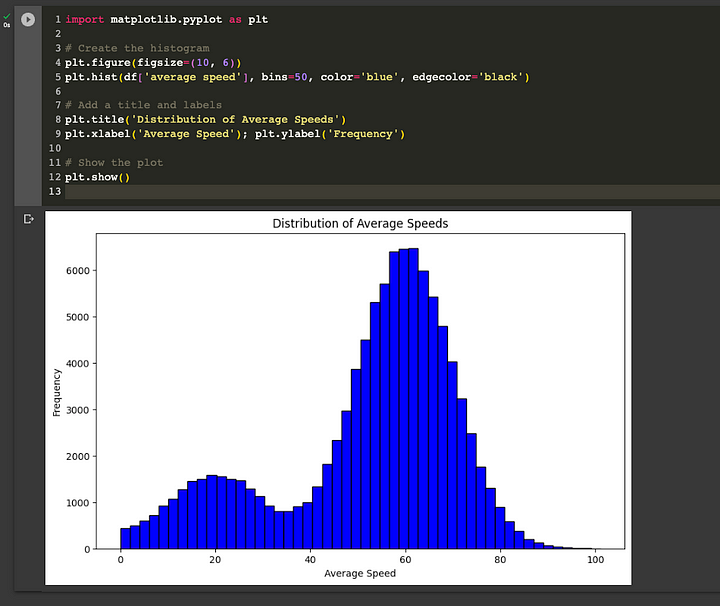}
        \caption{Histogram of average vehicle speeds obtained from the synthetic telematics dataset. This visualization is the result of running the Python code provided by ChatGPT, demonstrating the application of ChatGPT's generated code in analyzing synthetic data.}
        \label{fig:plot}
    \end{minipage}
\end{figure}

\subsection{Coherence Verification}
Synthetic data should adhere to certain logical principles. For instance, start times for trips must invariably precede the corresponding end times, and vehicular headings should remain plausible within the urban framework. A prompt to generate Python code for verifying this consistency could be phrased as follows: “What is the procedure to scrutinize timestamp data within a synthetic telematics dataset? Could a Python code snippet be crafted to ascertain if trip initiation and termination times align logically?”

\subsection{Optimizing Synthetic Dataset Assessment}
These meticulous steps facilitate a thorough assessment of the synthetic dataset's applicability and credibility, even when a real-world comparative dataset is not available. It's crucial to emphasize that the process of generating a synthetic dataset is iterative. Each evaluation serves to enhance understanding of the dataset's strong points and areas requiring improvement, thereby fostering the generation of progressively superior synthetic datasets.

\section{Conclusion}
The process of synthetic dataset creation, as outlined in this paper, involves multiple iterations and is heavily reliant on providing precise and comprehensive prompts to guide the AI model, in this instance, ChatGPT. The capacity to generate synthetic data presents transformative opportunities across various sectors, including telematics, where privacy concerns and data scarcity can obstruct progress and innovation.

This exploratory procedure highlighted the advantages of synthetic dataset creation, with emphasis on their potential to preserve data privacy, augment real-world data, and provide controlled environments for experimental purposes. Simultaneously, light was shed on potential difficulties, such as ensuring the quality and representativeness of synthetic data and the necessity for meticulous, iterative prompt development.

A practical demonstration navigated the generation of a synthetic telematics dataset for Columbus, Ohio. Beginning with a basic prompt, the complexity was gradually increased, with each stage providing learning opportunities for refining the prompts to achieve the desired output. This step-by-step approach led to the creation of a dataset with 100,000 rows, demonstrating ChatGPT's capability to handle complex tasks.

An assessment of the synthetic dataset's quality was also conducted, focusing on the uniqueness, diversity, and relevance of the data points. Descriptive statistics and visualizations were utilized to explore data characteristics and the interrelations among variables. Without a real-world comparative dataset, the importance of applying multiple testing techniques and verifying the logical coherence and relevance of the synthetic data was acknowledged.

In conclusion, synthetic datasets generated by models like ChatGPT indeed hold significant potential. However, it is important to remain objective, acknowledging that they are not perfect replacements for real-world data and have inherent limitations. Nonetheless, their potential applications in sectors such as telematics are wide-ranging, providing new avenues for research and experimentation. With careful crafting of prompts and stringent evaluation of outputs, AI-generated synthetic data can become a powerful instrument in a data scientist's toolkit.

\bibliographystyle{unsrtnat}  

 \end{document}